\documentclass[12pt]{article}
\usepackage{aas_macros}
\usepackage{times}
\usepackage{geometry}
\usepackage[utf8]{inputenc}
\usepackage{enumitem,amssymb}
\usepackage{ragged2e}
\usepackage{graphicx}
\usepackage{fancyhdr}
\usepackage{hyperref}
\usepackage{tabularx}
\usepackage[
    natbib=true,
    style=numeric,
    sorting=none
]{biblatex}
\usepackage{color, colortbl}
\newlist{thematic}{itemize}{8}
\setlist[thematic]{label=$\square$}
\usepackage{pifont}

\newcommand\farcs{\mbox{$.\!\!^{\prime\prime}$}}%

\begin{document}
\raggedright
\huge{
Decisively Demonstrating Roman CGI's TTR5 Requirement by Reimaging a Newly-Discovered Brown Dwarf Orbiting a Bright Accelerating Star}
\linebreak
\large

Thayne Currie$^{1,2}$,
Mona El Morsy$^{1}$, Brianna Lacy$^{3}$,  Masayuki Kuzuhara$^{4}$, Naoshi Murakami$^{4}$, Danielle Bovie$^{1}$

1 Department of Physics and Astronomy, University of Texas-San Antonio, San Antonio, TX, USA \\
2 Subaru Telescope, National Astronomical Observatory of Japan, 
Hilo, HI, USA\\
3 Department of Astronomy \& Astrophysics, University of California-Santa Cruz, Santa Cruz, CA, USA\\
4 National Astronomical Observatory of Japan, 2-21-1 Osawa, Mitaka, Tokyo 181-8588, Japan

\justify{
\textbf{Summary:} We propose Roman Coronagraph project HLC/575 nm observations of a newly-discovered brown dwarf (HIP 71618 B) from the Subaru/OASIS survey of young accelerating stars, which is supported by NASA headquarters with the directive to identify targets for the Roman Coronagraph that could fulfill TTR5 requirements and be observed during the technology demonstration phase.   The target and multiple bright PSF references are within/close to the Roman Continuous Viewing Zone.  A high SNR detection of this companion would singlehandedly fulfill TTR5 and would be the first optical detection of a companion at $<$10$^{-6}$ contrast.  

Roman CPP reference star vetting prioritizing stars that can be paired with HIP 71618 would aid the execution of a successful technology demonstration.  Additional similar targets may be discovered from OASIS over the next few years that could increase CGI scheduling flexibility and enhance its scientific and technical return.  A close collaborative partnership with the CPP team could ensure that they are schedulable.

\pagebreak
\pagebreak
\noindent \textbf{Type of observation:} \\$\boxtimes$   Technology Demonstration\\
$\square$ Scientific Exploration\\\\

\noindent \textbf{Scientific / Technical Keywords:}  
high contrast performance

\noindent \textbf{Required Detection Limit:}  
\begin{tabular}{| c | c | c | c | c |}
\hline
$\geq$10$^{-5}$ & 10$^{-6}$ & 10$^{-7}$ & 10$^{-8}$ & 10$^{-9}$ \\ \hline
 & & x & & \\ \hline
\end{tabular}

\vspace{0.5cm}
\textbf{Roman Coronagraph Observing Mode(s):} 

\begin{tabular}{| c | c | c | c | c |}
\hline
Band &   Mode & Mask Type & Coverage & Support \\ \hline \hline

1, 575 nm &   Narrow FoV & Hybrid Lyot & 360$^{\circ}$ & Required (Imaging) \\
 & Imaging &  &  &  \\ \hline 
\end{tabular}

\justify{
\begin{center}
\begin{tabular}{| c | c | c | c | c |}
\hline
Name &  host star & detection & separation (") & description of target \\
  & V mag. & limit & (or extent)  & \\ \hline \hline
  HIP 71618 & 5.3 & 10$^{-7}$ & 0.25-0.28 & self-luminous brown dwarf\\
  HD 120315 & 1.86 & -- & -- & PSF Reference 1\\
  HD 127762 & 3.02 & -- & -- & PSF Reference 2\\
  \hline
\end{tabular}
\end{center}

}
\pagebreak

\begin{figure}[!ht]
    \includegraphics[width=0.4\textwidth,clip]{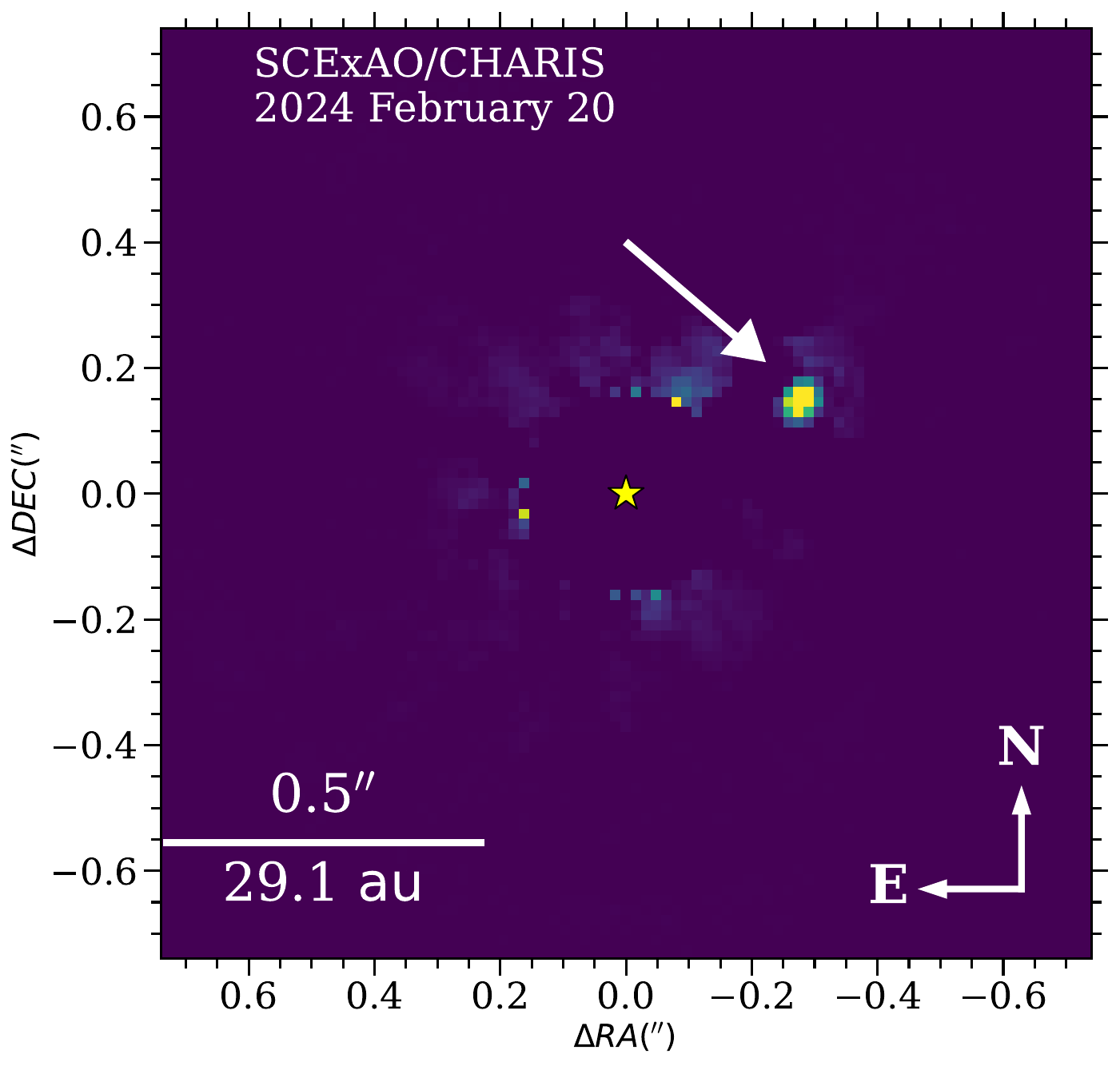}
     \includegraphics[width=0.6\textwidth,clip]{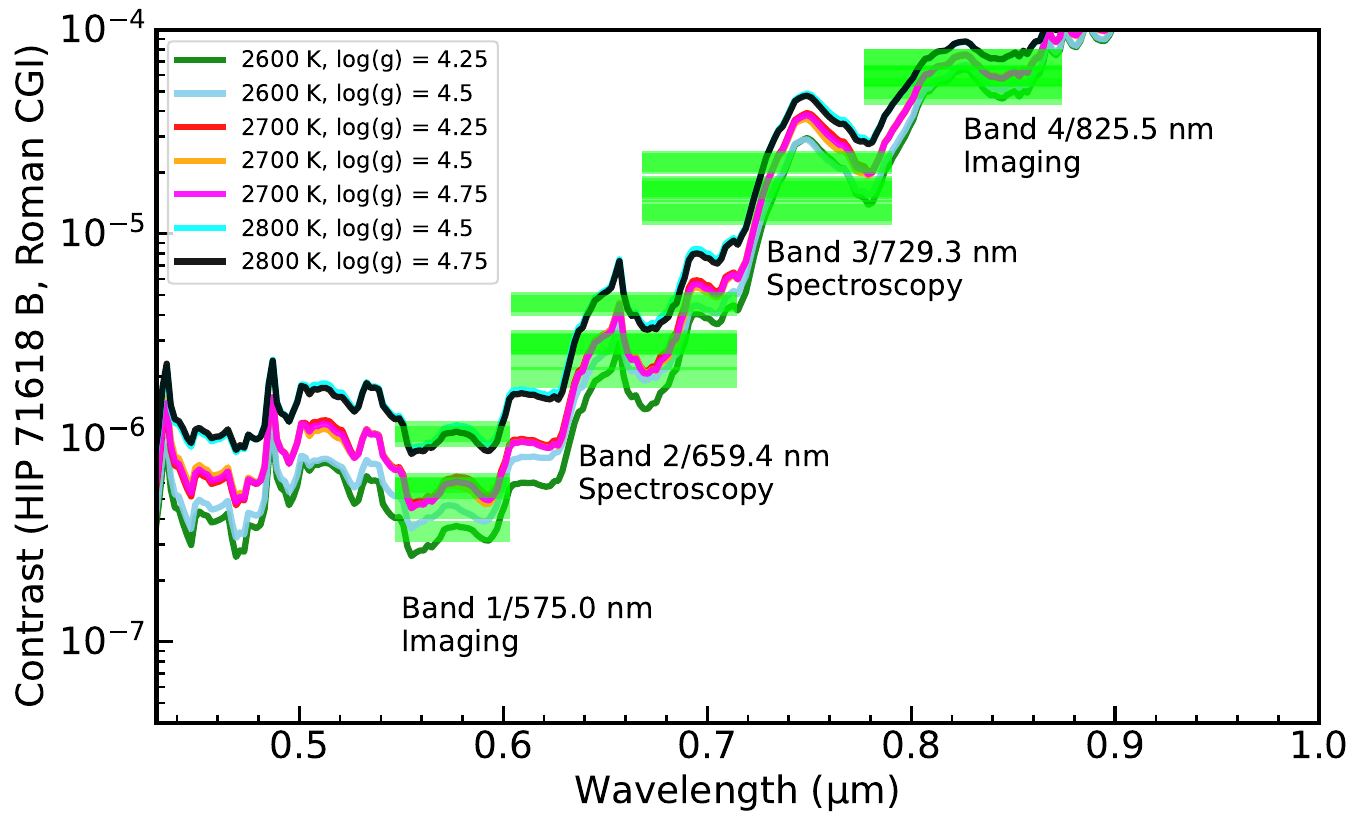}
    \vspace{-0.4in}
    \caption{(left) Discovery image of HIP 71618 B with SCExAO/CHARIS \citep{Jovanovic2015,Groff2016} at an angular separation of 0\farcs{}315 from the OASIS survey.   The target was selected by the same strategy used to discover and precisely characterize the HIP 99770 b planet and other planets/brown dwarfs by both imaging and astrometry \citep{Currie2023a,Kuzuhara2022,Franson2023,Tobin2024,Currie2025} 
    (right) HIP 71618 B's predicted contrast in the four main CGI passbands from updated versions of the \citeauthor{LacyBurrows2020} models.  Modeling its CHARIS spectra and NIRC2 photometry yield a temperature of $\sim$2700 $\pm$ 100 $K$.   At 575 nm, it then has a contrast of 5 [3--10] $\times$10$^{-7}$.  
    }
\end{figure}

\justify{
\textbf{\Large Anticipated Technology / Science Objectives:}
The Roman CGI technological demonstration has a single Threshold Technical Requirment (TTR5) for success: 
achieve a 5-$\sigma$ contrast better than    
10$^{-7}$ {at $\lambda_{\rm c}$ $\le$ 600 nm ($>$10\% bandpass) located 6--9 $\lambda$/D ($\sim$ 0\farcs{}3--0\farcs{}45) from a bright (V$_{\rm AB}$ $\le$ 5) star\footnote{By construction, demonstrating the 10$^{-7}$ contrast threshold in more challenging configurations -- e.g. on a star fainter than V = 5 or on a companion located slightly interior to 6 $\lambda$/D -- would also fulfill TTR5} within 10 hours.  Achieving TTR5 ``by analysis" of the residual noise around a companionless star near a PSF reference is a possibility.  However, as of Dec 10 2025 there is no public-facing program from the Roman Community Participation Program (CPP) to vet such stars nor an announcement to carry out such observations in the near future.

The best and far more striking alternative is to demonstrate TTR5 by reimaging a known, cool substellar companion found from ground-based infrared high-contrast imaging.
However, as has been pointed out many times before \citep[e.g.][]{ElMorsy2024}, prior to December 3 2025 \textbf{the peer-reviewed literature lacked any known imaged companion whose redetection with CGI Band 1 would demonstrably fulfill this requirement}\footnote{\citet{Hom2025} describe $\beta$ Pic, HD 206893, and 47 Uma as ``potential science targets" but did not yield contrast predictions with updated atmosphere models to show that they are in fact detectable at 575 nm.  A closer analysis shows that all three are not good technology demonstration targets.  E.g. $\beta$ Pic's bright debris disk impedes the detection of $\beta$ Pic b at 575 nm.  Other commonly mentioned image systems are likewise problematic.  Published models \citep{LacyBurrows2020} suggest that 51 Eri b is too faint, while updates to these models likewise show HR 8799 e to be undetectable \citep[see][]{ElMorsy2025}.}.  As explained in \citet{ElMorsy2025}, alternate plans -- demonstrating TTR5 with the detection of a mature radial-velocity-identified planet -- are extremely risky and likely impractical.

We propose to demonstrate TTR5 by reimaging HIP 71618 B \citep{ElMorsy2025}, a low-mass companion (likely a brown dwarf) newly discovered from the Observations of Accelerators with SCExAO Imaging Survey (OASIS) (PI T. Currie; \citealt{ElMorsy2024}) specifically funded by NASA Headquarters as \textit{Strategic Mission Support} to identify Roman Coronagraph targets that could fulfill TTR5 (Fig. 1) and be observable during the technology demonstration phase.  Analysis of SCExAO/CHARIS and Keck/NIRC2 astrometry \citep{Brandt2021} find HIP 71618 B at $\rho$ $\sim$ 0\farcs{}3, in an edge-on orbit.  Its predicted positions in late 2026 through 2028 place it within the CGI dark hole region where OS 11 predicts a flat contrast floor.  The star is within the Roman Continuous Viewing Zone \citep{Rose2023} and is within close proximity to at least two potential PSF reference stars: HD 120315 and HD 127762 (Fig. 2, left panel).   Based on updated models from \citeauthor{LacyBurrows2020}, its average predicted contrast at 575nm is $\sim$ 5 $\times$10$^{-7}$ ([3$\times$10$^{-7}$--10$^{-6}$]) (Fig. 1).  \textit{Thus, a high SNR detection of HIP 71618 B at 575 nm would demonstrate TTR5}.

This program provides a quick, direct, and efficient way to fulfill TTR5.  The companion is in an edge-on orbit, moving closer in projected separation with time. Thus, this observation is best executed as early as possible during the observation phase.

\begin{figure}[!h]
 \includegraphics[width=0.5\textwidth,clip]{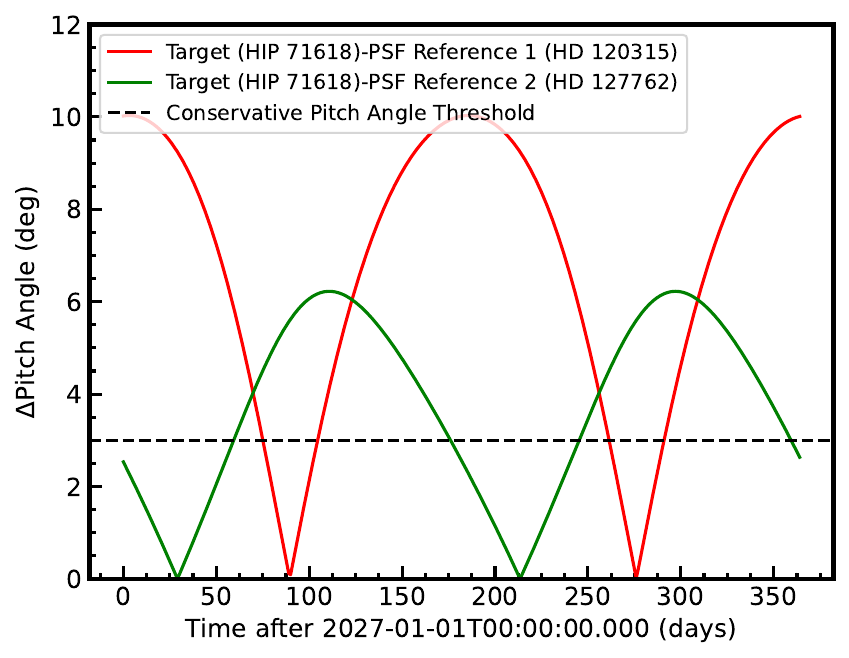}
      \includegraphics[width=0.5\textwidth,clip]{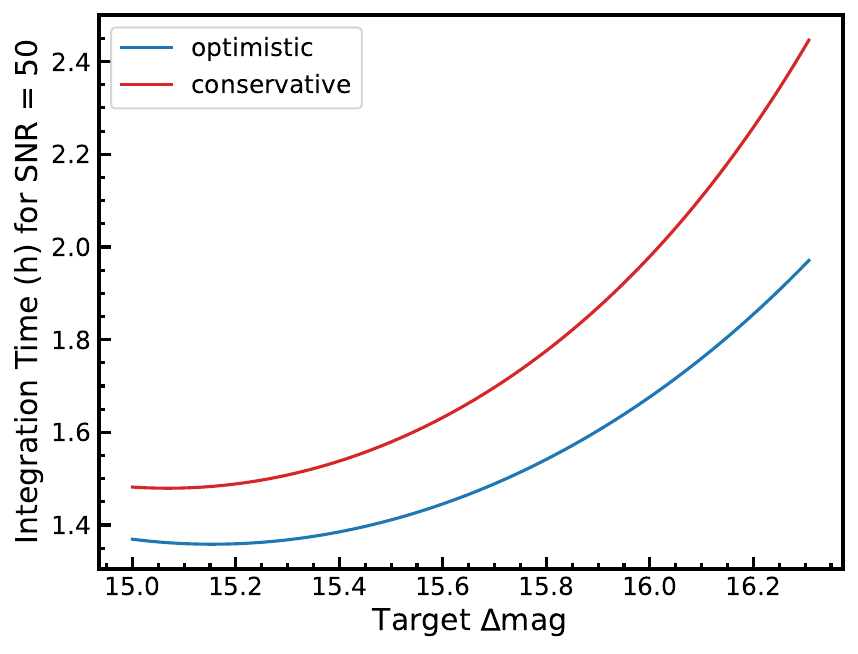}
    \vspace{-0.4in}
    \caption{(left) Relative Pitch Angle between HIP 71618 and two potential reference stars during the calendar year of 2027.  At least one of the PSF references has $\Delta$ Pitch Angle $\le$ 3$^{o}$ for over 90\% of the year.  Keepout maps (not shown) reveal that HIP 71618 and HD 120315 are continuously viewable, while HD 127762 has two small excluded periods due to a smaller solar angle.  (right) Integration time needed to achive a SNR = 50 detection of HIP 71618 B vs. HIP 71618 B contrast at 575 nm for conservative and optimistic CGI performances.  We assume a solar system exozodi level, though HIP 71618 lacks evidence for circumstellar dust.  }
\end{figure}

\textbf{Additional Preparatory Observations and Broader Outlook}:
The Roman CPP team's efforts to vet PSF reference stars at deep contrast is ongoing \citep[see ][]{Hom2025}.  Both shallow AO imaging and speckle interferometry fail to identify a companion to HD 120315 that would exclude it from being a suitable PSF reference star.  Archival Keck/NIRC2 imaging data likewise does not reveal a bright contaminating companion.   These data disfavor the presence of G, K, and early M companions but do not yet rule out the presence of lower-mass companions.   As both stars are too far north to be efficiently observed with SPHERE, deep extreme AO imaging with SCExAO/CHARIS is likely required to propertly vet these stars.

Finally, we stress that upcoming OASIS survey time (through Fall 2026 and possibly beyond) and later programs (e.g. similar to the proposed Subaru-Roman synergy) will likely identify additional companions suitable for CGI follow up to meet TTR5 and other Objectives, enhancing the instrument's scientific and technical return.   A close collaborative partnership with CPP team will ensure that these targets are schedulable. 

\textbf{Estimate of Time Needed}: We use the Roman Coronagraphic Instrument Exposure Time Calculator (Corgi-ETC) to estimate our required time allocation for optimistic and conservative scenarios for CGI's performance.  

For a contrast of 5$\times$10$^{-7}$, Corgi-ETC predicts on-source observing times of 1.5 hrs and 1.75 hrs to achieve 10$^{-7}$ 5-$\sigma$ contrasts for the optimistic and conservative performance scenarios, respectively.   For the worst-case scenario ($\Delta$575 nm = 3$\times$10$^{-7}$), these times increase to only 1.9 hrs and 2.5 hrs (Fig. 2, right panel). Adopting the CGI primer efficiency estimate (24 hours clock time for 14 hours observing time),  we then estimate a nominal clock time of 2.6 hrs and 3 hours for the optimistic and conservative scenarios with worst-case scenario values of 3.3 hrs and 4.3 hrs.  These values are well below the limits specified for TTR5 (10 hours).   


\printbibliography

@ARTICLE{Brandt2021,
       author = {{Brandt}, Timothy D. and {Dupuy}, Trent J. and {Li}, Yiting and {Brandt}, G. Mirek and {Zeng}, Yunlin and {Michalik}, Daniel and {Bardalez Gagliuffi}, Daniella C. and {Raposo-Pulido}, Virginia},
        title = "{orvara: An Efficient Code to Fit Orbits Using Radial Velocity, Absolute, and/or Relative Astrometry}",
      journal = {\aj},
         year = 2021,
        month = nov,
       volume = {162},
       number = {5},
          eid = {186},
        pages = {186},
          doi = {10.3847/1538-3881/ac042e},
archivePrefix = {arXiv},
       eprint = {2105.11671},
 primaryClass = {astro-ph.IM},
       adsurl = {https://ui.adsabs.harvard.edu/abs/2021AJ....162..186B}
}

@ARTICLE{Currie2023a,
       author = {{Currie}, Thayne and {Brandt}, G. Mirek and {Brandt}, Timothy D. and {Lacy}, Brianna and {Burrows}, Adam and {Guyon}, Olivier and {Tamura}, Motohide and {Liu}, Ranger Y. and {Sagynbayeva}, Sabina and {Tobin}, Taylor and {Chilcote}, Jeffrey and {Groff}, Tyler and {Marois}, Christian and {Thompson}, William and {Murphy}, Simon J. and {Kuzuhara}, Masayuki and {Lawson}, Kellen and {Lozi}, Julien and {Deo}, Vincent and {Vievard}, Sebastien and {Skaf}, Nour and {Uyama}, Taichi and {Jovanovic}, Nemanja and {Martinache}, Frantz and {Kasdin}, N. Jeremy and {Kudo}, Tomoyuki and {McElwain}, Michael and {Janson}, Markus and {Wisniewski}, John and {Hodapp}, Klaus and {Nishikawa}, Jun and {He{\l}miniak}, Krzysztof and {Kwon}, Jungmi and {Hayashi}, Masahiko},
        title = "{Direct imaging and astrometric detection of a gas giant planet orbiting an accelerating star}",
      journal = {Science},
         year = 2023,
        month = apr,
       volume = {380},
       number = {6641},
        pages = {198-203},
          doi = {10.1126/science.abo6192},
archivePrefix = {arXiv},
       eprint = {2212.00034},
 primaryClass = {astro-ph.EP},
       adsurl = {https://ui.adsabs.harvard.edu/abs/2023Sci...380..198C}
}

@ARTICLE{Currie2025,
       author = {{Currie}, Thayne and {Li}, Yiting and {El Morsy}, Mona and {Lacy}, Brianna and {Vincent}, Maria and {Tobin}, Taylor L. and {Kuzuhara}, Masayuki and {Chilcote}, Jeffrey and {Guyon}, Olivier and {Gu}, Ziying and {Bovie}, Danielle and {Peng}, Dillon and {An}, Qier and {Brandt}, Timothy D. and {De Rosa}, Robert J. and {Deo}, Vincent and {Groff}, Tyler D. and {Janson}, Markus and {Kasdin}, N. Jeremy and {Lozi}, Julien and {Marois}, Christian and {Mennesson}, Bertrand and {Murakami}, Naoshi and {Nielsen}, Eric and {Sagynbayeva}, Sabina and {Skaf}, Nour and {Thompson}, William and {Tamura}, Motohide and {Uyama}, Taichi and {Vievard}, S{\'e}bastien and {Zurlo}, Alice},
        title = "{SCExAO/CHARIS and Gaia Direct Imaging and Astrometric Discovery of a Superjovian Planet 3--4 lambda/D from the Accelerating Star HIP 54515}",
      journal = {arXiv e-prints},
         year = 2025,
        month = dec,
          eid = {arXiv:2512.02159},
        pages = {arXiv:2512.02159},
          doi = {10.48550/arXiv.2512.02159},
archivePrefix = {arXiv},
       eprint = {2512.02159},
 primaryClass = {astro-ph.EP},
       adsurl = {https://ui.adsabs.harvard.edu/abs/2025arXiv251202159C}
}

@INPROCEEDINGS{ElMorsy2024,
       author = {{El Morsy}, Mona and {Currie}, Thayne and {Kuzuhara}, Masayuki and {Chilcote}, Jeffrey and {Guyon}, Olivier and {Tobin}, Taylor L. and {Brandt}, Timothy and {An}, Qier and {Anh}, Kyohoon and {Bovie}, Danielle and {Deo}, Vincent and {Groff}, Tyler and {Gu}, Ziying and {Janson}, Markus and {Jovanovic}, Nemanja and {Li}, Yiting and {Lawson}, Kellen and {Lozi}, Julien and {Lucas}, Miles and {Marois}, Christian and {Murakami}, Naoshi and {Nielsen}, Eric L. and {Norris}, Barnaby and {Skaf}, Nour and {Tamura}, Motohide and {Thomson}, William and {Uyama}, Taichi and {Vievard}, Sebastien},
        title = "{Design, scientific goals, and performance of the SCExAO survey for planets around accelerating stars}",
    booktitle = {Adaptive Optics Systems IX},
         year = 2024,
       editor = {{Jackson}, Kathryn J. and {Schmidt}, Dirk and {Vernet}, Elise},
       series = {Society of Photo-Optical Instrumentation Engineers (SPIE) Conference Series},
       volume = {13097},
        month = aug,
          eid = {130977I},
        pages = {130977I},
          doi = {10.1117/12.3019513},
       adsurl = {https://ui.adsabs.harvard.edu/abs/2024SPIE13097E..7IE}
}

@ARTICLE{ElMorsy2025,
       author = {{El Morsy}, Mona and {Currie}, Thayne and {Lacy}, Brianna and {Tobin}, Taylor L. and {An}, Qier and {Li}, Yiting and {Gu}, Ziying and {Kuzuhara}, Masayuki and {Bovie}, Danielle and {Peng}, Dillon and {Chilcote}, Jeffrey and {Guyon}, Olivier and {Brandt}, Timothy D. and {De Rosa}, Robert J. and {Deo}, Vincent and {Groff}, Tyler D. and {Janson}, Markus and {Kasdin}, N. Jeremy and {Lozi}, Julien and {Marois}, Christian and {Mennesson}, Bertrand and {Murakami}, Naoshi and {Nielsen}, Eric and {Sagynbayeva}, Sabina and {Skaf}, Nour and {Thompson}, William and {Tamura}, Motohide and {Uyama}, Taichi and {Vievard}, S{\'e}bastien and {Zurlo}, Alice},
        title = "{OASIS Survey Direct Imaging and Astrometric Discovery of HIP 71618 B: A Substellar Companion Suitable for the Roman Coronagraph Technology Demonstration}",
      journal = {arXiv e-prints},
         year = 2025,
        month = dec,
          eid = {arXiv:2512.02126},
        pages = {arXiv:2512.02126},
          doi = {10.48550/arXiv.2512.02126},
archivePrefix = {arXiv},
       eprint = {2512.02126},
 primaryClass = {astro-ph.SR},
       adsurl = {https://ui.adsabs.harvard.edu/abs/2025arXiv251202126E}
}

@ARTICLE{Franson2023,
       author = {{Franson}, Kyle and {Bowler}, Brendan P. and {Bonavita}, Mariangela and {Brandt}, Timothy D. and {Chen}, Minghan and {Samland}, Matthias and {Zhang}, Zhoujian and {Lueber}, Anna and {Heng}, Kevin and {Kitzmann}, Daniel and {Wolf}, Trevor and {Jones}, Brandon A. and {Tran}, Quang H. and {Bardalez Gagliuffi}, Daniella C. and {Biller}, Beth and {Chilcote}, Jeffrey and {Crepp}, Justin R. and {Dupuy}, Trent J. and {Faherty}, Jacqueline and {Fontanive}, Cl{\'e}mence and {Groff}, Tyler D. and {Gratton}, Raffaele and {Guyon}, Olivier and {Jensen-Clem}, Rebecca and {Jovanovic}, Nemanja and {Kasdin}, N. Jeremy and {Lozi}, Julien and {Magnier}, Eugene A. and {Mu{\v{z}}i{\'c}}, Koraljka and {Sanghi}, Aniket and {Theissen}, Christopher A.},
        title = "{Astrometric Accelerations as Dynamical Beacons: Discovery and Characterization of HIP 21152 B, the First T-dwarf Companion in the Hyades}",
      journal = {\aj},
         year = 2023,
        month = feb,
       volume = {165},
       number = {2},
          eid = {39},
        pages = {39},
          doi = {10.3847/1538-3881/aca408},
archivePrefix = {arXiv},
       eprint = {2211.09840},
 primaryClass = {astro-ph.SR},
       adsurl = {https://ui.adsabs.harvard.edu/abs/2023AJ....165...39F}
}

@INPROCEEDINGS{Groff2016,
       author = {{Groff}, Tyler D. and {Chilcote}, Jeffrey and {Kasdin}, N. Jeremy and
         {Galvin}, Michael and {Loomis}, Craig and {Carr}, Michael A. and {Brand
        t}, Timothy and {Knapp}, Gillian and {Limbach}, Mary Anne and
         {Guyon}, Olivier and {Jovanovic}, Nemanja and {McElwain}, Michael W. and
         {Takato}, Naruhisa and {Hayashi}, Masahiko},
        title = "{Laboratory testing and performance verification of the CHARIS integral field spectrograph}",
    booktitle = {Ground-based and Airborne Instrumentation for Astronomy VI},
         year = "2016",
       series = {Society of Photo-Optical Instrumentation Engineers (SPIE) Conference Series},
       volume = {9908},
        month = "Aug",
          eid = {99080O},
        pages = {99080O},
          doi = {10.1117/12.2233447},
       adsurl = {https://ui.adsabs.harvard.edu/abs/2016SPIE.9908E..0OG}
}

@ARTICLE{Hom2025,
       author = {{Hom}, Justin and {Wolff}, Schuyler G. and {Clark}, Catherine A. and {Ciardi}, David R. and {Deveny}, Sarah J. and {Howell}, Steve B. and {Greenbaum}, Alexandra Z. and {Littlefield}, Colin and {Anche}, Ramya M. and {Bailey}, Vanessa P. and {Brandner}, Wolfgang and {Chauvin}, Ga{\"e}l and {Girard}, Julien H. and {Kern}, Brian and {Mamajek}, Eric and {Mennesson}, Bertrand and {Savransky}, Dmitry and {Stapelfeldt}, Karl R. and {Biller}, Beth A. and {Brinjikji}, Marah and {Kuzuhara}, Masayuki and {Millar-Blanchaer}, Maxwell A. and {Mizuki}, Toshiyuki and {Schragal}, Nicholas T. and {Vega-Pallauta}, Macarena C. and {Wang}, Jason J. and {De Rosa}, Robert J. and {Douglas}, Ewan S. and {Macintosh}, Bruce and {Zhang}, Jingwen and {Coronagraph Community Participation Program}, the Roman},
        title = "{CoronaGraph Instrument Reference stars for Exoplanets (CorGI-REx) I. Preliminary Vetting and Implications for the Roman Coronagraph and Habitable Worlds Observatory}",
      journal = {arXiv e-prints},
         year = 2025,
        month = nov,
          eid = {arXiv:2511.08862},
        pages = {arXiv:2511.08862},
          doi = {10.48550/arXiv.2511.08862},
archivePrefix = {arXiv},
       eprint = {2511.08862},
 primaryClass = {astro-ph.SR},
       adsurl = {https://ui.adsabs.harvard.edu/abs/2025arXiv251108862H}
}

@ARTICLE{Jovanovic2015,
       author = {{Jovanovic}, N. and {Martinache}, F. and {Guyon}, O. and {Clergeon}, C. and
         {Singh}, G. and {Kudo}, T. and {Garrel}, V. and {Newman}, K. and
         {Doughty}, D. and {Lozi}, J. and {Males}, J. and {Minowa}, Y. and
         {Hayano}, Y. and {Takato}, N. and {Morino}, J. and {Kuhn}, J. and
         {Serabyn}, E. and {Norris}, B. and {Tuthill}, P. and {Schworer}, G. and
         {Stewart}, P. and {Close}, L. and {Huby}, E. and {Perrin}, G. and
         {Lacour}, S. and {Gauchet}, L. and {Vievard}, S. and {Murakami}, N. and
         {Oshiyama}, F. and {Baba}, N. and {Matsuo}, T. and {Nishikawa}, J. and
         {Tamura}, M. and {Lai}, O. and {Marchis}, F. and {Duchene}, G. and
         {Kotani}, T. and {Woillez}, J.},
        title = "{The Subaru Coronagraphic Extreme Adaptive Optics System: Enabling High-Contrast Imaging on Solar-System Scales}",
      journal = {\pasp},
         year = 2015,
        month = sep,
       volume = {127},
       number = {955},
        pages = {890},
          doi = {10.1086/682989},
archivePrefix = {arXiv},
       eprint = {1507.00017},
 primaryClass = {astro-ph.IM},
       adsurl = {https://ui.adsabs.harvard.edu/abs/2015PASP..127..890J}
}

\end{document}